\documentclass[11pt]{article} 
\usepackage{amsfonts,amscd,amsmath,amssymb,graphicx,color}

\newcommand{\oh}{\frac{1}{2}}
\def\m2{m^2_0}
\def\ep{\text{e}}
\def\a{\alpha'}
\def\zo{z_{\text{\tiny 0}}}
\def\g{\mathfrak{g}}
\def\s{\mathfrak{s}}
\def\xp{x_+}
\def\zt{z_{\text{\tiny T}}}

\def\zc{z_{\text{\tiny c}}}
\def\zt{z_{\text{\tiny T}}}

\title{Nonlocal and Mixed Condensates Tests of Gauge/String Duality}
\author{Oleg Andreev\thanks{Also at Landau Institute for Theoretical Physics, Moscow.}
\\ \\
{\it Arnold Sommerfeld Center for Theoretical Physics, LMU-M\"unchen,} \\
{\it Theresienstrasse 37, 80333 M\"unchen, Germany}}
\date{}

\begin{document} 

\vspace{-8cm} 
\maketitle 
\begin{abstract} 
Using a duality approach we give an example of modeling nonlocal and mixed condensates whose behavior mimics that of QCD. Quantitatively, the estimated value of the condensate parameter $\m2$ is approximately $0.6-0.7\,\text{GeV}^2$.
 \end{abstract}

\vspace{-10cm}
\begin{flushright}
LMU-ASC 63/10
\end{flushright}
\vspace{9cm}


\section{Introduction}

The determination of the condensates of the QCD vacuum is a very important issue in phenomenology of strong interactions \cite{svz}. As known, it can only be done in a non-perturbative formulation of the theory. At present, not much is known about the condensates from first principles. Some attempts exist to determine them on the lattice or in the models of the instanton vacuum. In practice, however, the QCD sum rules still remain a basic tool for doing so.\footnote{For a review and list of references, see \cite{review}.} Thus, there is a strong need for new alternative approaches to the problem.

The AdS/CFT correspondence \cite{malda} has opened new avenues for studying strongly coupled gauge theories. Although the original proposal was for conformal theories, various modifications have been found that produce, in particular, gauge/string duals with a mass gap, confinement, and supersymmetry breaking \cite{rev-ads}.

In this paper we continue our study of the condensates within gauge/string duality. In \cite{az4}, we estimated the value of the gluon 
condensate with a result that is surprisingly close to the original phenomenological estimate of \cite{svz}. As known, QCD is a very rich theory supposed to describe the whole spectrum of strong interaction phenomena. The question naturally arises: How well does gauge/string duality describe other condensates? Here, we attempt to {\it analytically} determine a function $Q$ and a parameter $\m2$, which appear in nonlocal and mixed condensates, as 
an important step toward answering this question. 

The simplest quark-gluon mixed condensate is that associated with a dimension-5 operator constructed from the quark and gluon fields as $\bar q\sigma^{\mu\nu}G_{\mu\nu}q$, where $\sigma^{\mu\nu}$ is an antisymmetric combination of $\gamma$-matrices. It is well known that it plays an important role in various QCD sum rules \cite{review}. 

The parameter $\m2$ appears as a constant of proportionality in the conventional parametrization \cite{review,rad-rev}

\begin{equation}\label{m20}
	\langle g\bar q\sigma_{\mu\nu}G^{\mu\nu} q\rangle=\m2\langle \bar qq\rangle
	\,,
\end{equation}
where $\langle\bar qq\rangle$ is a quark (chiral) condensate and $g$ is a gauge coupling constant.

One possible way to determine both the mixed condensate and the parameter $\m2$ is to consider a nonperturbative gauge invariant correlator, also 
known as the nonlocal condensate,\footnote{The literature on the nonlocal condensates is vast. For more details, see, e.g., \cite{rad-rev} and references therein.}

\begin{equation}\label{q2}
\Psi(x_1,x_2)=\langle\bar q(x_1)U_P(x_1,x_2)q(x_2)\rangle
\,,
\end{equation}
where $U_P(x_1,x_2)$ is a path-ordered Wilson line defined as $U_P(x_1,x_2)=P\exp\bigl[ig\int_0^1ds \frac{dx^\mu}{ds}\,A_\mu(x(s)) \bigr]$. Here $s$ is a parameter of the path running from $0$ at $x=x_1$ to $1$ at $x=x_2$. The path is taken to be a straight line. If one sets \cite{rad-rev}

\begin{equation}\label{Q-par}
\Psi(x_1,x_2)\equiv\langle \bar qq\rangle Q(r)\,,
\end{equation}
then $\m2$ is given by the coefficient of $r^2$, with $r=\vert x_1-x_2\vert$, in the expansion of the function $Q$ as $r\rightarrow 0$ 

\begin{equation}\label{q21}
	Q(r)=1-\frac{1}{16}\m2 r^2+O(r^4)
	\,.
\end{equation}
Note that this formula holds in Euclidean space. In Minkowski space it is modified by replacing $r^2\rightarrow -r^2$ \cite{rad-rev}. 

Before proceeding to the detailed analysis, let us set the basic framework for the dual description. We take the 
dual string spacetime as a product of five-dimensional space, with the Euclidean metric \footnote{This is the ansatz. At this time we do not know equations that provide such a solution. So, we follow "the inverse scattering problem": first, we suggest a solution, then we look for its phenomenological relevance.}

\begin{equation}\label{metric}
	ds^2={\cal G}_{nm}dX^n dX^m=
R^2\frac{h}{z^2}
\left(dt^2+d\vec{x}^2+dz^2\right)
\,,\qquad
h=\ep^{\s z^2}
\,
\end{equation}
and some five-dimensional internal space $X$. Here $\s$ is a parameter whose value can be fixed from the heavy quark potential (Cornell model) or the slope of the Regge trajectory of $\rho(n)$ mesons. We also take a constant dilaton and discard other background fields. In what follows, 
we assume a trivial dependence on the internal space $X$. Unlike other string duals, this model does share a few key features with QCD that singles it out and makes it very attractive for phenomenology: First, the model is a nearly conformal theory at UV, where it leads to the quadratic corrections \cite{q2,nz}. Second, the model results in a phenomenologically satisfactory description of the confining potential \cite{az1,gian}. Finally, at finite chemical potential its extension provides the phenomenologically acceptable equation of state for cold quark matter \cite{cold}. 


\section{Calculating the Correlator}

To begin with, we will make an ansatz for computing the function $Q$ within gauge/string duality whose justification, initially, is that it 
combines the ingredients at hand in the most natural way. Gradually, further evidence for the ansatz will emerge. 

Let us set the quark operators and the Wilson line on a four-manifold which is the boundary of a five-dimensional manifold. We will assume that the function $Q$ is given in terms of the area (in string units) of a surface in the five-dimensional manifold by \footnote{We discuss some issues 
that arise in attempting to include a constant proportionality below.}

\begin{equation}\label{Q}
	Q=\ep^{-S}
	\,.
\end{equation}
Here $S$ is the area of a surface sketched in Figure 1.

\begin{figure}[ht]
\centering
\includegraphics[width=5.7cm]{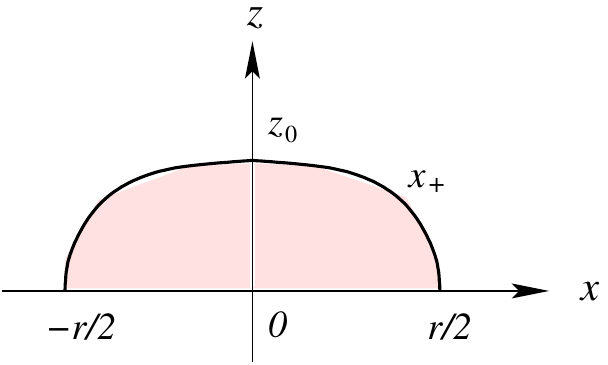}
\caption{\small{A surface in a five-dimensional manifold. The boundary is at $z=0$. 
The surface is bounded by a curved profile of a static string stretched between the quark sources set 
at $x=\pm r/2$ and a straight Wilson line along the $x$-axis.}} 
\end{figure}

It is worth noting that a similar representation suggested in \cite{p-loop} has proven successful for studying the expectation value of the 
Polyakov loop in pure gauge theories. Recently, it was shown in \cite{F2} that it also reproduces the exponential decay of the correlator 
$\Psi$ at large separations, as expected in QCD. In the model with dynamical quarks it appeared in \cite{kutasov}. A difference is that here the Wilson loop goes along an internal direction on a five-dimensional boundary. 

\subsection{Shape of Static String}

Following \cite{az1}, we will now describe the shape of the static string stretched between the quark sources in the background geometry \eqref{metric}. First we set the quark operators at $x=\pm r/2$ on the boundary ($z=0$), as shown in Figure 1. Next, we make use of the Nambu-Goto action 
endowed with the background metric

\begin{equation}\label{NG}
	S=\frac{1}{2\pi\a}\int d^2\xi\sqrt{\det{\cal G}_{nm}\partial_\alpha X^n\partial_\beta X^m}
\,.
\end{equation}

In the static gauge $\xi^1=t$ and $\xi^2=x$, the action becomes 

\begin{equation}\label{ng2}
	S=\frac{\g}{2\pi}T\int ^{r/2}_{-(r/2)}dx\,\frac{h}{z^2}\sqrt{1+(z')^2}
	\,,
\end{equation}
where $\g=R^2/\a$. A prime denotes a derivative with respect to $x$.

Then, as in \cite{az1}, we find the equation of motion for $z$

\begin{equation}\label{eom}
	zz''+2(1-\s z^2)(1+(z')^2)=0
	\,
\end{equation}
as well as its first integral

\begin{equation}\label{integral}
	\frac{h}{z^2\sqrt{1+(z')^2}}=\text{const}
	\,.
\end{equation}
The integration constant can be expressed in terms of the maximum value of $z$. On symmetry grounds, the function $z(x)$ has a maximum at 
$x=0$. It is $z(0)=\zo$, as shown in Figure 1.

Since $z(x)$ is an even function, we pick a fundamental domain defined by $0\leq x\leq r/2$. Moreover, it is convenient to use the inverse function $\xp (z)$ subject to 
the boundary conditions

\begin{equation}\label{bc}
	\xp(0)=\oh r\,,\quad
	\xp(\zo)=0
	\,.
\end{equation}

Using these boundary conditions, the solution to equation \eqref{integral} in the fundamental domain is 

\begin{equation}\label{xp}
	\xp (z)=\sqrt{\frac{\lambda}{\s}}\int^1_{z\sqrt{\frac{\s}{\lambda}}} du\,u^2\ep^{\lambda (1-u^2)}(1-u^4\ep^{2\lambda(1-u^2)})^{-(1/2)}	
	\,,
\end{equation}
where $\lambda=\s\zo^2$. Finally, the solution on the interval $-r/2\leq x\leq 0$ is simply $x_-(z)=-\xp (z)$.

If we set $z=0$, then \eqref{xp} becomes

\begin{equation}\label{r}
	r=2\sqrt{\frac{\lambda}{\s}}\int^1_0 du\,u^2\ep^{\lambda (1-u^2)}(1-u^4\ep^{2\lambda(1-u^2)})^{-(1/2)}
\,,
\end{equation}
which is the form in which it is written in \cite{az1}.

\subsection{Renormalized Area}
Having found the function $\xp (z)$ describing the string shape, we can now calculate the renormalized area of the surface shown in Figure 1.

To begin with, we choose the gauge $\xi^1=x$ and $\xi^2=z$. Then we substitute this into the action \eqref{NG} to obtain

\begin{equation}\label{NG2}
	S=\frac{\g}{\pi}\sqrt{\frac{\s}{\lambda}}\int_0^1 du\,\frac{\ep^{\lambda u^2}}{u^2}\xp 
	\,.
\end{equation}
Here we have rescaled $z$ such that $z=\zo u$.

The integral is divergent at $u=0$ due to the factor $z^{-2}$ in the metric \eqref{metric}. To proceed, therefore, we need to regularize it. A way 
to deal with this divergence is to cut off the integral at $u=\epsilon$. So, we have (up to terms vanishing for $\epsilon\rightarrow 0$)

\begin{equation}\label{NG3}
S=\frac{\g}{\pi\zo}
\biggl[
\frac{r}{2}\Bigl(\frac{1}{\epsilon}-1\Bigr)+
\int^1_0 \frac{du}{u^2}\ep^{\lambda u^2}\Bigl(\xp -\frac{r}{2}\Bigr)+
\frac{r}{2}\int^1_0 \frac{du}{u^2}\Bigl(\ep^{\lambda u^2}-1\Bigr)
\biggr]
\,.
\end{equation}

Now, by integration by parts, the action is

\begin{equation}\label{NG4}
S=\frac{\g}{\pi}
\biggl[\oh\frac{r}{\zo\epsilon}+
\int^1_0 du\,u^2\ep^{\lambda (1-u^2)}
\bigl(1-u^4\ep^{2\lambda(1-u^2)}\bigr)^{-(1/2)}
\bigl(\sqrt{\pi\lambda}\,\text{Erfi}(\sqrt{\lambda}u)-u^{-1}\ep^{\lambda u^2}\bigr)
\biggr]
\,,
\end{equation}
where $\text{Erfi}(u)$ is the imaginary error function.

As in \cite{F2}, we use the modified minimal subtraction scheme to deal with the power divergence. So, we subtract 
$\frac{\g\,r}{2\pi\zo}(\frac{1}{\epsilon}-c)$, where $c$ denotes a constant whose value must be specified from renormalization 
conditions. Finally, using \eqref{r} we get

\begin{equation}\label{NG5}
S=\frac{\g}{\pi}\int^1_0 du\,u^2\ep^{\lambda (1-u^2)}
\bigl(1-u^4\ep^{2\lambda(1-u^2)}\bigr)^{-(1/2)}
\bigl(\sqrt{\pi\lambda}\text{Erfi}(\sqrt{\lambda}u)-u^{-1}\ep^{\lambda u^2}+c\bigr)
\,.
\end{equation}

\subsection{Q and $\m2$}

Actually, the function $Q=\ep^{-S}$ is written in parametric form given by equations \eqref{r} and \eqref{NG5}. At this time it is not clear to us how to eliminate the parameter $\lambda$ and find $Q$ as a function of $r$. We can, however, gain some important insights into the problem by considering two limiting cases.

First, let us have a close look at $r(\lambda)$. According to \cite{az1}, it is a continuously growing function defined on the 
interval $[0,1]$. The asymptotic behavior near zero is given by 

\begin{equation}\label{rs}
	r=\frac{1}{\rho}\sqrt{\frac{\lambda}{\s}}\Bigl(1-\oh\lambda(1-\pi\rho^2)+O(\lambda^2)\Bigr)
	\,,
\end{equation}
where $\rho=\Gamma^2(\tfrac{1}{4})/(2\pi)^{3/2}$. Since $\lambda\rightarrow 0$ means $r\rightarrow 0$, small $\lambda$'s correspond to small values of 
$r$.

The asymptotic behavior near $1$ is given by 

\begin{equation}\label{rl}
	r=-\frac{1}{\sqrt{\s}}\ln(1-\lambda) +O(1)
	\,.
\end{equation}
Thus, this region corresponds to large values of $r$.

At this point a comment is in order. If we take $\lambda\in [0,1]$, then there exists the upper bound on the maximum value of $z$ such that 
$\zo\leq\zc=1/\sqrt{\s}$ \cite{az1}. In view of the formula \eqref{rl}, this means that $\zo\rightarrow\zc$ as $r\rightarrow\infty$.

Once the behavior of $r(\lambda)$ is understood, we can use it to study the properties of the function $Q$ at short and long distances. 

We begin with the case of small $r$. Expanding the right hand side of equation \eqref{NG5} in $\lambda$, in next-to-leading order we have

\begin{equation}\label{Ss}
	S=\frac{\g}{\pi}\Bigl(-\frac{\pi}{4}+\frac{c}{2\rho}-\frac{c}{4\rho}\lambda(1-\pi\rho^2)+O(\lambda^2)\Bigr)
	\,.
\end{equation}
Then combining this with \eqref{rs}, we find the desired behavior of the function $Q$ at short distances 

\begin{equation}\label{Qs}
	Q=Q_0\Bigl(1-\frac{1}{16}\m2 r^2+O(r^4)\Bigr)
	\,,
\end{equation}
where 
\begin{equation}\label{m2}
	\m2=4c\g\,\s\rho \left(\rho^2-\pi^{-1}\right)
	\
\end{equation}
and 
\begin{equation}\label{N}
	Q_0=\exp\Bigl\{\frac{\g}{4}\Bigl(1-\frac{2c}{\pi\rho}\Bigr)\Bigr\}
	\,.
\end{equation}

In a similar spirit, we can explore the long distance behavior of $Q$. It follows from \eqref{NG5} that in the neighborhood of $\lambda=1$ the renormalized area behaves as 

\begin{equation}\label{Sl}
	S=-\frac{\g}{2\pi}\bigl(\sqrt{\pi}\text{Erfi}(1)-\ep+c\bigr)\ln(1-\lambda)+O(1)
	\,.
\end{equation}
Along with the relation \eqref{rl}, this means that the function $Q$ decays exponentially at long distances as

\begin{equation}\label{lQ}
	Q\simeq\ep^{-Mr}
	\,,
\end{equation}
where \footnote{The length scale $\xi=M^{-1}$ known as the correlation length has been also computed in \cite{F2}.}
\begin{equation}\label{M}
	M=\frac{\g\sqrt{\s}}{2\pi}\Bigl(\sqrt{\pi}\,\text{Erfi}(1)-\ep +c\Bigr)
	\,.
\end{equation}
This behavior is precisely analogous to what is expected in QCD.

Having understood the two limiting cases, we are now able to make some estimates relevant to phenomenology. 

To make an estimate of the parameter $\m2$, we need to fix a value for the constant $c$. First, we will fix it 
from the standard normalization of $Q$ that is $Q(0)=1$ \cite{rad-rev}.\footnote{While it may seem natural to assume this normalization in our framework, where it is equivalent to the fact that the surface shown in Figure 1 shrinks to a point as $r\rightarrow 0$, it requires a caveat. In section 3, we 
will propose a more refined way.} From \eqref{N}, we find that $c=\pi\rho/2\approx 1.3$. Given the value of $c$, we can estimate the parameter $\m2$, with the result

\begin{equation}\label{m2-est}
	\m2=2\g\,\s\rho^2\left(\pi\rho^2-1\right)\approx 0.70\,\,\text{GeV}^2
	\,.
\end{equation}
Here we have used $\s\approx 0.45\,\,\text{GeV}^2$ and $\g\approx 0.94$ as it follows from the fits to the slopes for the Regge trajectory of $\rho(n)$ mesons \cite{q2} and the linear term of the Cornell potential \cite{az1}. 

There is a long history of attempts to estimate the value of $\m2$ \cite{review}. According to the original phenomenological estimate based on the QCD sum rules \cite{ioffe}, which is widely accepted, it is given by $\m2=0.8\pm 0.2\,\,\text{GeV}^2$. Thus, our estimate is very satisfying at this point. Of course, there are other estimates which are also close to the original value. For comparison, lattice simulations \cite{lattice} and the field correlator method \cite{simonov} both result in somewhat larger values, of order $1\,\,\text{GeV}^2$.\footnote{At this point, we are somewhat formal and compare 
the results in different renormalization schemes.}

To complete the picture, let us estimate the parameter $M$ resulting from long distances. With our values for the parameters, we have

\begin{equation}\label{M-est}
	M\approx 0.15\,\,\text{GeV}
\,.
\end{equation}
This value is surprisingly closed to the pion mass that may be an indication that at long distances the correlator is dominated by the lightest 
meson contribution. 

On the other hand, our stringy construction suggests a natural normalization condition $M=\sqrt{\sigma}$, where $\sigma$ is the string tension (the coefficient of the linear term in the Cornell potential). For a typical value of $\sigma=0.18\,\,\text{GeV}^2$, it gives $M\approx 0.42\,\,\text{GeV}$ together with $c\approx 4.0$. This simple estimate allows us to draw the following conclusions: 

(i) The value of $M$ is now close to the lowest energy level of the heavy-light mesons in the heavy quark effective theory \cite{rad-rev} which is around $0.45\,\,\text{GeV}$. This suggests that at long distances the correlator is dominated by this lowest state. 

(ii) $\m2$ is definitely larger than $0.8\,\,\text{GeV}^2$ (up to a factor of $2.6$), and that it is of order $2.1\,\,\text{GeV}^2$. Interestingly, the lattice calculation of \cite{lattice2} and the instanton liquid model of \cite{shuryak}  
yield even larger values, about $2.4-2.5\,\,\text{GeV}^2$. 

(iii) An additional free parameter is needed if one also wants $Q(0)=1$. A simple way to introduce it is to use a slightly modified ansatz
$Q=\ep^{-S}/Q_0$, where $Q_0$ is the normalization constant defined by eq.\eqref{N}.

Some comments about the value of $\m2$. The procedure for determining $\m2$ is very sensitive to the procedure of extracting the power divergence. This is consistent with the literature, where the values show significant scatter (up to a factor of $3$). One of the motivations for this work was to calculate it in a new nonperturbative approach that would shed light on this problem.

Finally, let us present the results of numerical calculations. The parametric equations \eqref{r} and \eqref{NG5} predict a characteristic form for 
the function $Q$, as shown by the upper curve in Figure 2.  

\begin{figure}[ht]
\centering
\includegraphics[width=6.8cm]{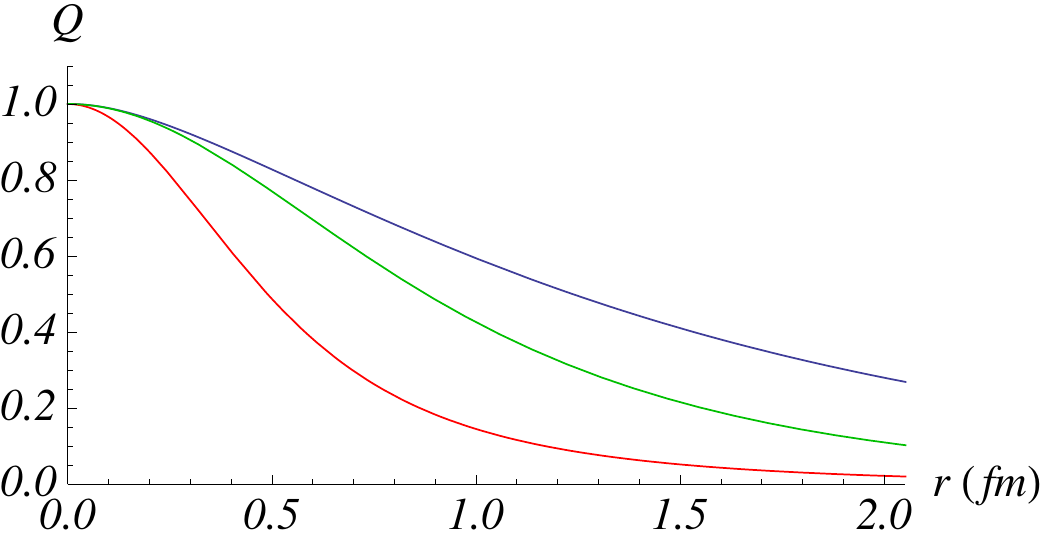}
\caption{\small{Q as a function of $r$. The upper blue curve comes from the parametric equations \eqref{r} and \eqref{NG5}. The lower red and green curves correspond to the models A and B, respectively. For all the cases, $\m2=0.7\,\text{GeV}^2$ and $M=0.15\,\text{GeV}$.}} 
\end{figure}

It is of great interest to compare this form with phenomenological models. Various possibilities of those have been discussed in the 
literature. Here we will consider two simple models, say A and B, for the vacuum distribution function $f$ which in Euclidean space is related to $Q$ by an integral transform $Q(r)=\int_0^\infty ds\,\ep^{-sr^2/4}f(s)$. Following \cite{rad-rev}, we take $f_A(s)=A\exp\{-M^2/s-s^2/a\}$ and 
$f_B(s)=B\exp\{-M^2/s-s/b\}$. Both these functions reproduce $Q\sim\ep^{-Mr}$ for large $r$. The normalization is chosen so that the zeroth moment is $\int_0^\infty ds\,f(s)=1$ or, equivalently, $Q(0)=1$. In addition, the first moment must obey $\int_0^\infty ds\,sf(s)=\m2/4$. We see that in the 
phenomenologically important interval $0.1\,\text{fm}\lesssim r\lesssim 1.5\,\text{fm}$ our model is closer to the model B whose distribution function 
falls exponentially at large $s$. For $r\leq 1\,\text{fm}$ the agreement between the models is quite good. The maximum discrepancy occurred at $r=1\,\text{fm}$ is of order 25 \%.

\section{Many Open Problems}

There is a large number of open problems associated with the circle of ideas explored in this paper. In this section we list a few.

(i) Our analysis is not very accurate for various reasons. For one thing, the string breaking effect must be included in the calculation of the 
renormalized area $S$. What happens when a light $\bar qq$ pair is created? The surface is now modified by the string decay, like that of Figure 3. 
Thus, one might expect that this would require improvement of the analysis of section 2 at large distances. At this point it is worth noting that the string breaking scale is of order $1.1-1.2\,\text{fm}$. Therefore, it is tempting to say that the discrepancy with the phenomenological models of Figure 2 for $r\gtrsim 1.2\,\text{fm}$ is a result of the string breaking effect.

\begin{figure}[ht]
\centering
\includegraphics[width=5.6cm]{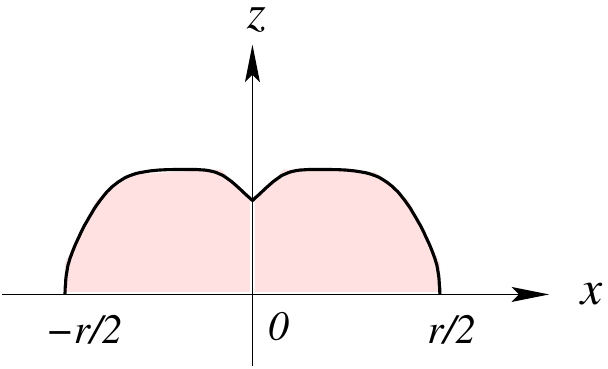}
\caption{\small{A surface modified by a light pair creation, there appears a cusp on the curve at $x=0$.}} 
\end{figure}

Also, the normalization of the nonperturbative correlator $\Psi$ (function $Q$) at $r=0$ is problematic: one might worry that this is a 
territory of perturbative QCD. It is therefore interesting to see what happens if the correlator is normalized at a length scale $r_0\not=0$. For doing so, let us set $r_0=0.2\,\text{fm}$ and see whether $\m2$ has a reasonable value. As before, we take $\s\approx 0.45\,\text{GeV}^2$ and $\g\approx 0.94$. Using \eqref{r}, we find numerically that a solution to $r(\lambda)=r_0$ is $\lambda\approx 0.23$. 
With this value of $\lambda$, the renormalized area \eqref{NG5} vanishes at $c\approx 1.13$. Finally, we get from \eqref{m2}

\begin{equation}\label{m2-1GeV}
	\m2\approx 0.60\,\,\text{GeV}^2
\,,
\end{equation}
with $Q(0.2\,\text{fm})=1$. This is still a satisfying and reasonable value. However, our analysis doesn't involve stringy quasiclassical corrections 
which may be important.\footnote{Although we have no satisfactory explanation of why these corrections are small, it is interesting to note that the classical string computation of the heavy quark potential shows a remarkable agreement with the lattice data \cite{az1,white}.}

(ii) In the case of $\text{AdS}_5$ the parametric equations become trivial 

\begin{equation}\label{ads5}
	r=\frac{1}{\rho}\zo\,,\quad
	S=\frac{\g}{2\pi}\left(-\frac{\pi}{2}+\frac{c}{\rho}\right)
	\,.
\end{equation}
Note that $S$ is independent of $r$, as required by conformal invariance.

Now, a problem arises. If one tries to compute the function $Q$ within the model based on a truncated AdS space, where the shape of 
a static string is the same as in AdS space until the string is long enough to reach the cutoff (IR brane location) \cite{braga}, how can the expansion \eqref{q21} 
occur? \footnote{In this model the computation of the gluon condensate along the lines of \cite{az4} is also puzzling.}

(iii) An interesting observation of lattice simulations \cite{latticeT} is that at low temperatures (below $T_c$) the parameter $\m2$ is almost independent of the temperature. We can gain some understanding of this by writing the dual string spacetime either as \eqref{metric}, with $t$ a 
periodic variable of period $1/T$, or as \cite{az2}

\begin{equation}\label{metricT}
		ds^2=R^2\frac{h}{z^2}
	\left(fdt^2+d\vec{x}^2+f^{-1}dz^2\right)
	\,,\qquad
	h=\ep^{\s z^2}
	\,,\qquad
	f=1-\Bigl(\frac{z}{\zt}\Bigr)^4
	\,,
\end{equation}
where $\zt=1/(\pi T)$. Since small $r$'s correspond to small $\zo$'s, we expand the five-dimensional metric around the $\text{AdS}_5$ background by writing $h$ and $f$ as power series in $z$. The first term in the expansion \eqref{q21} is a constant which equals $1$. It comes from the $\text{AdS}_5$ metric. The second term determines the parameter $\m2$. It is due to the leading correction to the $\text{AdS}_5$ metric. In our case the correction is independent of the temperature, so the parameter $\m2$ is independent, too. This reasoning, however, requires a caveat at high temperatures, where 
the expansion around $\text{AdS}_5$ is no more appropriate. 
 
(iv) In section 2 we cut off the integral over $u$. What if we had chosen to cut off the integral over $z$ and then subtract $\frac{\g \,r}{2\pi}\left(\frac{1}{\epsilon}-c\right)$? In that case the large $r$ behavior of $S$ remains unaffected since $\zo\rightarrow 1/\sqrt{\s}$ as 
$r\rightarrow\infty$, but $\zo$ depends on $r$ at smaller $r$ so the small $r$ behavior of $S$ will be different.\footnote{This is also the case for 
the heavy quark potential computed in \cite{az1}. So it provides another reason for us to fix the constant $\g$ at large distances.} Although we have no 
satisfactory explanation of why the regularization of section 2 is more appropriate, we chose it because it leads to quite reasonable results.

(v) As a final remark, we should point out that in holographic QCD $\m2$ is treated as a free parameter whose value is 
fixed by phenomenology \cite{kim}. In such an approach one starts from a five-dimensional effective field theory action, somehow motivated by string 
theory but without higher derivatives terms (stringy $\alpha'$ corrections), and tries to fit it to QCD as much as possible. One might think of 
criticizing this approach on the grounds that it doesn't include (all) stringy $\alpha'$ corrections to the effective action (for example, see \cite{reece}). Therefore, it is tempting to determine the value of $\m2$ by saying that it can be done as a result of the $\alpha'$ corrections. As we have seen above, the effective string theory description provides evidence for this by showing the $\alpha'$-dependence of $\m2$.

\vspace{.25cm}
{\bf Acknowledgments}

\vspace{.25cm}
This work was supported in part by DFG within the Emmy-Noether-Program under Grant No.HA 3448/3-1 and the Alexander von Humboldt Foundation under Grant No.PHYS0167. We would like to thank S. Hofmann, V.I. Zakharov, and especially P. Weisz for discussions and comments. Finally, we wish to express our gratitude to the Aspen Center for Physics for warm hospitality while this work was in progress.


\end{document}